# CORRELATIONS BETWEEN STRUCTURE AND DYNAMICS IN COMPLEX NETWORKS

Luciano da Fontoura Costa[1,*], Olaf Sporns[2], Lucas Antiqueira[3],
Maria das Graças V. Nunes[3], Osvaldo N. Oliveira Jr.[1]

[1] Instituto de Física de São Carlos, Universidade de São Paulo, Brazil
[2] Department of Psychological and Brain Sciences, Indiana University, USA
[3] Instituto de Ciências Matemáticas e de Computação, Universidade de São Paulo, Brazil

**Previous efforts in complex networks research focused mainly on the topological features of such networks, but now also encompass the dynamics. In this Letter we discuss the relationship between structure and dynamics, with an emphasis on identifying whether a topological hub, i.e. a node with high degree or strength, is also a dynamical hub, i.e. a node with high activity. We employ random walk dynamics and establish the necessary conditions for a network to be topologically and dynamically fully correlated, with topological hubs that are also highly active. Zipf's law is then shown to be a reflection of the match between structure and dynamics in a fully correlated network, as well as a consequence of the rich-get-richer evolution inherent in scale-free networks. We also examine a number of real networks for correlations between topology and dynamics and find that many of them are not fully correlated.**

We address the relationship between structure and dynamics of complex networks by taking the steady-state distribution of the frequency of visits to nodes – a dynamical feature – obtained by performing random walks[1] along the networks. A complex network[2-5] is taken as a graph with directed edges with associated weights, which are represented in terms of the weight matrix $W$. The $N$ nodes in the network are numbered as $i = 1, 2, \ldots, N$ and a directed edge with weight $M$, extending from node $j$ to node $i$, is represented as $W(i, j) = M$. No self-connections (loops) are considered. The *in-* and *outstrength* of a node $i$ – abbreviated as $os(i)$ and $is(i)$, correspond to the sum of the weights of the in- and outbound connections, respectively. The stochastic matrix $S$ for such a network is

$$S(i,j) = W(i,j) / os(j) \qquad (1)$$

---

[*] Corresponding Author: P.O. Box 369, 13560-970, São Carlos, SP, Brazil. Tel. +55 16 3373 9858; Fax +55 16 3371 3616. E-mail luciano@if.sc.usp.br.

The matrix $S$ is assumed irreducible, i.e. any of its nodes can be accessible from any other node, which allows the definition of a unique and stable steady state. An agent, placed at any initial node $j$, chooses among the adjacent outbound edges of node j with probability equal to $S(i, j)$. This step is repeated a large number of times $T$, and the frequency of visits to each node $i$ is calculated as $v(i) =$ (number of visits during the walk)$/T$. In the steady state (i.e. after a long time period $T$), $\vec{v} = S\vec{v}$ and the frequency of visits to each node along the random walk may be calculated in terms of the eigenvector associated to the unit eigenvalue. For proper statistical normalization we set $\sum_p v(p) = 1$. The dominant eigenvector of the stochastic matrix has theoretically and experimentally been verified to be remarkably similar to the corresponding eigenvector of the weight matrix, implying that the adopted random walk model shares several features with other types of dynamics, including linear and non-linear summation of activations and flow in networks.

With the frequency of visits to nodes (i.e. the 'activity' of the nodes) obtained as above, the correlation between activity and topology can be quantified in terms of the Pearson correlation coefficient, $r$. For full correlation, i.e. $r = 1$, two conditions must be jointly satisfied: (i) the network must be completely connected, i.e. $S$ is irreducible, and (ii) for any node, the instrength must be equal to the outstrength.

The proof of the statement above is as follows. Because the network is fully connected, its stochastic matrix $S$ has a unit eigenvector in the steady state, i.e. $\vec{v} = S\vec{v}$. As $S(i, j) = W(i, j)/os(j)$, the $i$-th element of the vector $S\vec{os}$ is given as

$$
\begin{aligned}
&S(i,1)os(1) + S(i,2)os(2) + \cdots + S(i,N)os(N) = \\
&\frac{W(i,1)}{os(1)}os(1) + \frac{W(i,2)}{os(2)}os(2) + \cdots + \frac{W(i,N)}{os(N)}os(N) = \\
&W(i,1) + W(i,2) + \cdots + W(i,N) = is(i)
\end{aligned}
\quad (2)
$$

Since, by hypothesis, $is(i) = os(i)$ for any $i$, both $\vec{os}$ and $\vec{is}$ are eigenvectors of $S$ associated to the unit eigenvalue, and therefore $\vec{os} = \vec{is} = \vec{v}$, implying full correlation between frequency of visits and corresponding node strengths.

An implication of this derivation is that for perfectly correlated networks, the frequency of symbols produced by random walks will be equal to the outstrength/instrength distributions. Therefore, an outstrength scale-free[3] network must produce sequences obeying Zipf's law[6], and vice versa. If, on the other hand, the node distribution is Gaussian, the frequency of visits to nodes will also be a Gaussian function; that is to say, the distribution of nodes is replicated in the node activation. A fully correlated network will have $r = 1$. This is illustrated in Fig. 1a and b, which show $r = 1$ for a text by Darwin[7] and the network of airports in the USA[8]. Zipf's law is known to apply to the former type of networks[9].

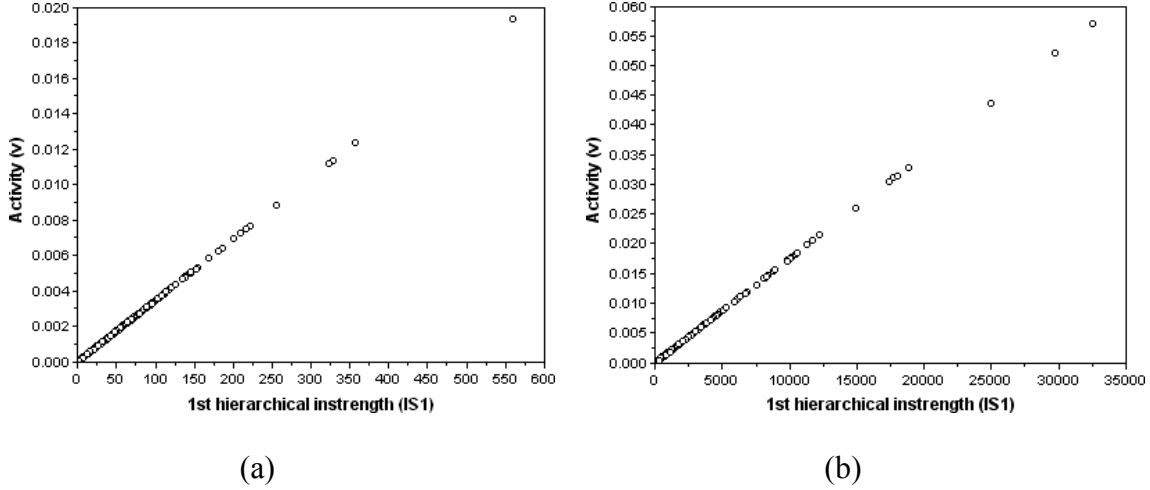

(a)　　　　　　　　　　　　　　　(b)

Fig. 1 – The activity (frequency of visits) of each node in terms of its instrength (also called cumulative 1st hierarchical instrength, as explained below) for the networks obtained from Darwin's text (a) and American airports (b). The word association network was obtained by representing each distinct word as a node while the edges were established by the sequence of immediately adjacent words in the texts after the removal of stopwords[10] and lemmatization[11]. More specifically, the fact that word $U$ has been followed by word $V$ $M$ times during the text is represented as $W(V,U) = M$. The airports network presents a link between two airports if there exists at least one flight between them. The number of flights performed in one month was used as the strength of the edges.

We obtained $r$ for various real networks (Table 1), including the fully correlated networks mentioned above. To interpret these data, we recall that a small $r$ means that a hub (large in- or outstrength) in topology is not necessarily a center of activity. Notably, in all cases considered $r$ is greater for the in- than for the outstrength. This may be understood with a trivial example of a node from which a high number of links emerge (implying large outstrength) but which has only very few inbound links. This node, in a random walk model, will be rarely occupied and thus cannot be a center of activity, though it will strongly affect the rest of the network by sending activation to many other targets. Understanding why a hub in terms of instrength may fail to be very active is more subtle. Consider a central node receiving links from many other nodes arranged in a circle, i.e. the central node has a large instrength, but with the surrounding nodes possessing small instrength. In other words, if a node $i$ receives several links from nodes with low activity, this node $i$ will likewise be fairly inactive. In order to further analyze the latter case, we may examine the correlations between the frequency of visits to each node $i$ and the *cumulative hierarchical in- and outstrengths* of that node. The hierarchical degrees[12-14] of a network node provide a natural extension of the traditional concept of node degree. The cumulative hierarchical outstrength of a node $i$ at the hierarchical level $h$ corresponds to the sum of the weights of the edges extending from the hierarchical level $h$ to the subsequent level $h+1$, plus the outstrengths obtained from hierarchy 1 to $h$-1. Similarly, the cumulative instrength of a node $i$ at hierarchical level $h$ is the sum of the weights of the edges from hierarchical level $h+1$ to the previous level $h$, plus the instrengths obtained from hierarchy 1 to $h$-1. The traditional in- and outstrength are the cumulative hierarchical in- and outstrength at hierarchical level 1 (see Supplementary Methods for a more detailed definition of cumulative hierarchical degree). Because complex networks are often also small world structures, it suffices to consider hierarchies up to 2 or 3 edges.

For the least correlated network analyzed, viz. that of the largest connected cluster in the network of WWW links between the pages contained in the massey.ac.nz domain (Massey University - New Zealand)[15, 16] – Fig. 2, activity could not be related to instrength at any hierarchical level. Because the Pearson coefficient corresponds to a single real value, it cannot adequately express the co-existence of the many relationships between activity and degrees present in this specific network, as well as possibly heterogeneous topologies. Very similar results were obtained for other WWW network, which indicates that the reasons why topological hubs have not been highly active cannot be identified at the present moment (see, however, discussion for higher correlated networks below).

Table 1 – Number of nodes (#Nodes), number of edges (#Edges), means and standard deviations of the clustering coefficient (CC), cumulative hierarchical instrengths for levels 1 to 4 (IS1 to IS4), cumulative hierarchical outstrengths for levels 1 to 4 (OS1 to OS4), and the Pearson correlation coefficient between the activation and all cumulative hierarchical instrengths and outstrengths ($C_{IS1}$ to $C_{OS4}$) for the complex networks considered in the present work.

|  | Cortex | *C. elegans* | Airports | Darwin | Wodehouse | WWW |
|---|---|---|---|---|---|---|
| **#Nodes** | 53 | 191 | 280 | 3678 | 3705 | 10810 |
| **#Edges** | 826 | 2449 | 4160 | 22095 | 16939 | 158102 |
| **CC** | 0.60 ± 0.15 | 0.22 ± 0.11 | 0.62 ± 0.41 | 0.04 ± 0.11 | 0.03 ± 0.08 | 0.60 ± 0.21 |
| **IS1** | 25.89 ± 9.42 | 100.82 ± 110.03 | 2041.07 ± 4323.33 | 7.87 ± 22.15 | 5.29 ± 16.15 | 14.63 ± 155.87 |
| **IS2** | 217.13 ± 56.68 | 1183.32 ± 960.60 | 76068.88 ± 53936.38 | 329.61 ± 648.33 | 188.45 ± 385.21 | 176.00 ± 917.67 |
| **IS3** | 285.02 ± 27.13 | 3543.97 ± 1118.85 | 110381.09 ± 35614.97 | 3352.93 ± 2716.07 | 1977.58 ± 1758.30 | 879.71 ± 2635.18 |
| **IS4** | 285.68 ± 27.13 | 4164.04 ± 535.73 | 113662.07 ± 32404.79 | 6943.53 ± 2470.62 | 4830.73 ± 1876.14 | 2468.12 ± 4528.49 |
| **OS1** | 25.89 ± 11.87 | 100.82 ± 73.69 | 2041.07 ± 4329.44 | 7.87 ± 22.15 | 5.29 ± 16.15 | 14.63 ± 10.58 |
| **OS2** | 217.96 ± 89.94 | 1156.76 ± 675.14 | 76049.93 ± 54196.34 | 313.16 ± 626.72 | 187.60 ± 394.19 | 176.00 ± 131.02 |
| **OS3** | 296.98 ± 34.93 | 3071.82 ± 806.15 | 110771.60 ± 35721.52 | 3234.23 ± 2705.50 | 1961.32 ± 1778.45 | 913.55 ± 495.34 |
| **OS4** | 298.94 ± 32.19 | 3532.41 ± 473.59 | 114054.35 ± 32493.50 | 6753.76 ± 2454.90 | 4823.73 ± 1853.97 | 2356.92 ± 1200.37 |
| $C_{IS1}$ | 0.83 | 0.78 | 1.00 | 1.00 | 1.00 | 0.15 |
| $C_{IS2}$ | 0.58 | 0.84 | 0.33 | 0.86 | 0.82 | 0.09 |
| $C_{IS3}$ | 0.24 | 0.43 | 0.11 | 0.42 | 0.43 | 0.13 |
| $C_{IS4}$ | 0.24 | 0.35 | 0.08 | 0.20 | 0.22 | 0.11 |
| $C_{OS1}$ | 0.39 | 0.20 | 1.00 | 1.00 | 1.00 | 0.00 |
| $C_{OS2}$ | 0.30 | 0.01 | 0.33 | 0.87 | 0.81 | -0.03 |
| $C_{OS3}$ | -0.03 | -0.19 | 0.11 | 0.42 | 0.43 | -0.05 |
| $C_{OS4}$ | -0.07 | -0.33 | 0.07 | 0.20 | 0.22 | -0.07 |

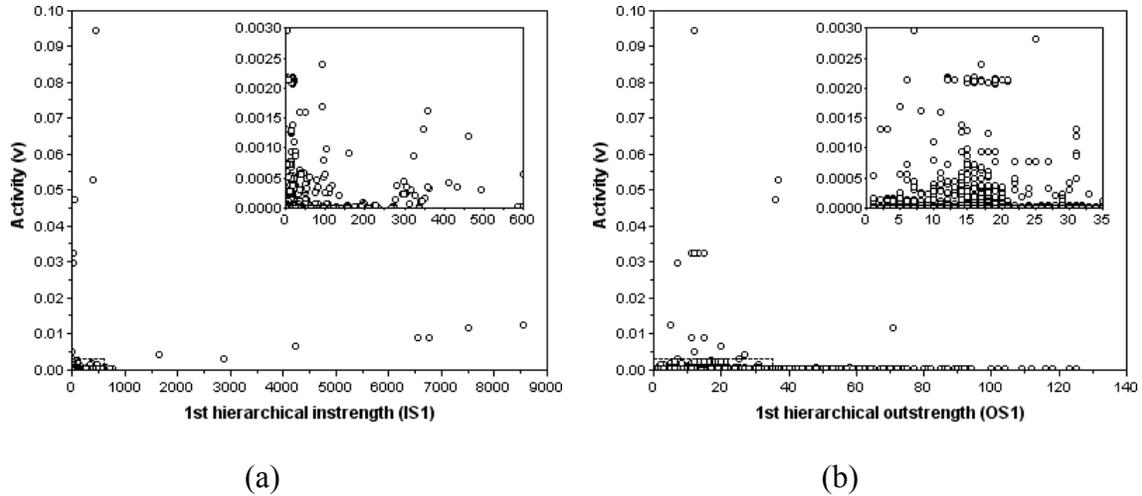

(a)                                 (b)

Fig. 2 – Analysis of the correlation between activity and node instrength for the WWW links between New Zealand domains. No clear correlation exists with respect to the cumulative 1st hierarchical in- (a) or outstrengths (b). Such a result is a consequence of the great intricacy of this large network, which involves several correlation substructures, which cannot be expressed into the single Pearson correlation coefficient.

However, for the two neuronal structures of Table 1 that are not fully correlated, activity was shown to increase with the cumulative 1st and 2nd hierarchical instrengths, as illustrated in comparing Fig. 3a and 3b for the network defined by the interconnectivity between cortical regions of the cat[17]. Similar results were obtained for the network of synaptic connections in *C. Elegans*[18]. In the cat cortical network, each cortical region is represented as a node, and the interconnections are reflected by the network edges. Significantly, in a previous paper[19], it was shown that when connections between cortex and thalamus were included, the correlation between activity and outdegree increased significantly. This could be interpreted as a result of increased efficiency with the topological hubs becoming highly active.

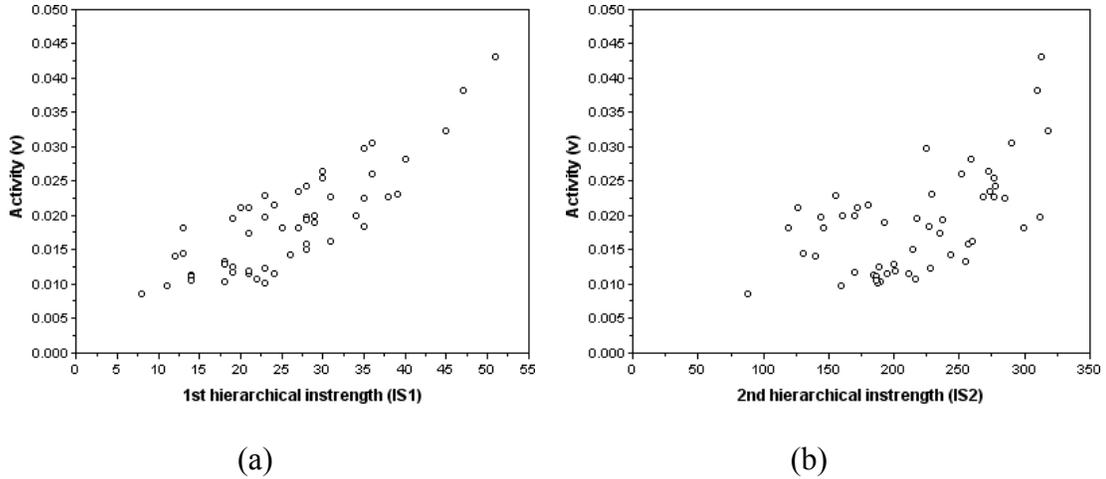

(a)                          (b)

Fig. 3 – The correlations between the frequency of visits to nodes vs. the cumulative $1^{st}$ hierarchical instrength (a) and the cumulative $2^{nd}$ hierarchical instrength (b) obtained from the cat cortical network.

Because of the many factors that may cause hubs failing to be active, one may gain insights into the importance of cumulative hierarchical instrengths by considering the model shown in Fig. 4a. Here, a star-shaped subnetwork (left-hand side of the figure) has been randomly attached, through $P$ incoming edges, to a larger random network containing 50 nodes and 250 directed edges (right-hand side of the figure). Note that the cumulative hierarchical indegree of node $i$ at hierarchy 1 is equal to $M$ (number of incoming neighbors), and its cumulative hierarchical indegree at hierarchy 2 is equal to $M + P$. We quantified the activity at node $i$ in terms of several values of $P$ for 100 different realizations (i.e. varying the target of the interconnections), obtaining the graph shown in Fig. 4b. The graph points to two regimes: an brief initial portion observed for small values of $P$, indicating strong positive correlation between the activity and cumulative hierarchical indegree, followed by a region of saturation, where the correlation is greatly diminished. The region with small correlation grows when the number of connections $P$ is increased relatively to the number of nodes in the larger network. This construction involving two attached networks has also been considered for studying the effect of varying levels of clustering coefficients of the reference node *i*. The results, also obtained after 100 realizations of each case, are shown in the Fig. 4c, from which it is clear that the increase of the clustering coefficient reduces the relative activation of the reference node, as the random walk agent will spend more time going from each of the neighbors. Therefore, both the cumulative hierarchical degree structure and the clustering coefficient contribute to changing the correlation between activation and node degree in subtle ways.

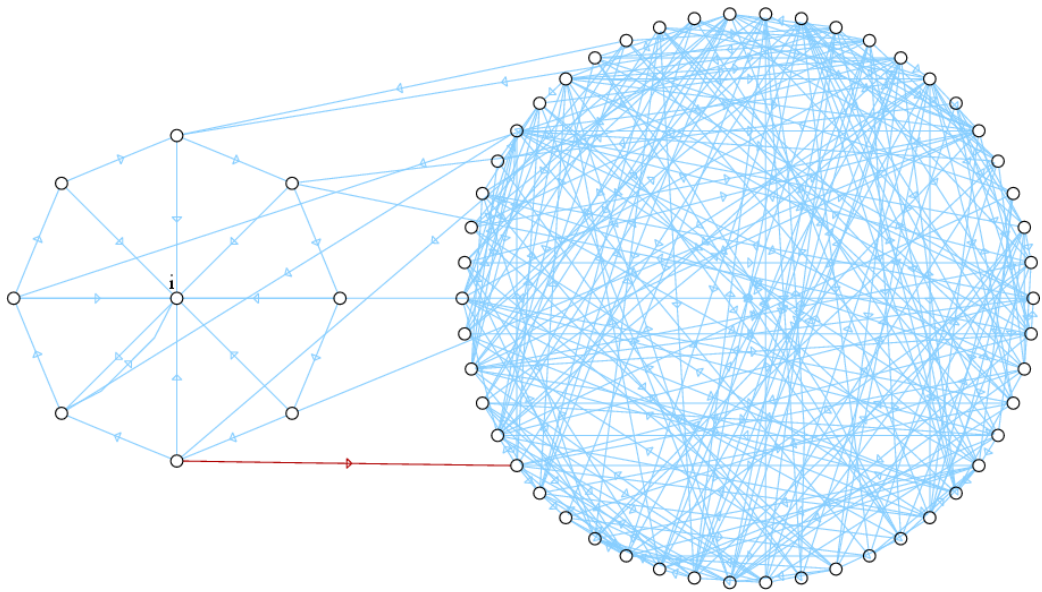

(a)

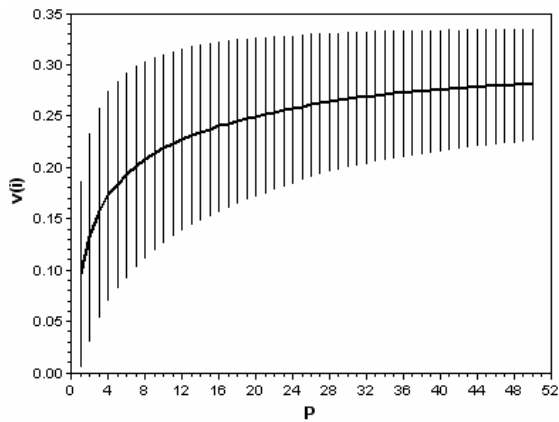

(b)

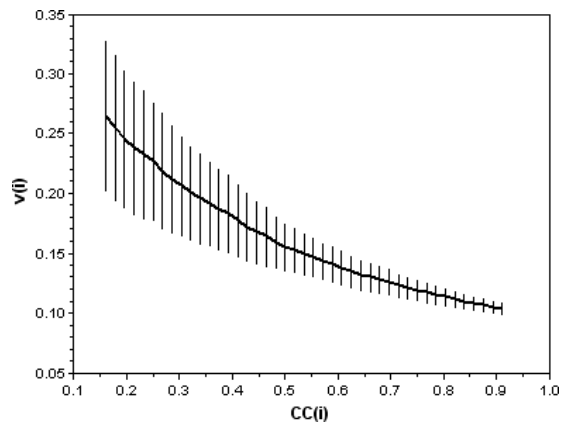

(c)

Fig. 4 – The correlation between the activity of node $i$ and its cumulative $2^{nd}$ hierarchical indegree can be investigated by randomly attaching a subnetwork such as that in the left-hand portion of the structure in (a) to a substantially larger network (right-hand side). The two networks are connected through $P$ directed edges which come from the right-hand side (9 in the case of this example, plus one edge in the opposite direction, to ensure a connected network), plus a single edge extending from left to right (in red). The curve quantifying the activity in terms of the cumulative $1^{st}$ hierarchical degree for node $i$, considering 100 realizations for each connectivity, is shown in (b), being characterized by two regimes. For $P < 10$, there is high positive correlation, while the correlation is diminished for $P > 10$. The effect of increasing clustering coefficients on reducing the activation of the reference node is shown in (c).

This model shows that a node linked to a set of highly connected nodes with large cumulative hierarchical instrengths, is likely to be more active than another node with the same instrength (cumulative $1^{st}$ hierarchical instrength) but lower cumulative $2^{nd}$ hierarchical instrength, as indicated by the results of the neuronal networks already discussed. Nevertheless, a more thorough analysis of such dependencies should consider the activation flow between nodes. Furthermore, for the fully correlated networks, such as word associations obtained for texts by Darwin[7] and Wodehouse[20], activity increased basically with the square of the cumulative $2^{nd}$ hierarchical instrength (see Supplementary Fig. 2). In addition, the correlations obtained for these two authors are markedly distinct, as the Wodehouse work is characterized by substantially steeper increase of frequency of visits for large instrength values (see Supplementary Fig. 3). Therefore, the results considering higher cumulative hierarchical degrees may serve as a feature for authorship identification.

In conclusion, we have established a set of conditions for full correlation between topological and dynamical features of complex networks, and demonstrated that Zipf´s law can be naturally derived for fully correlated networks. In the cases where the network is not fully correlated, the Pearson coefficient may be used as a characterizing parameter. For a network with very small correlation, such as the WWW links between the pages in a New Zealand domain analyzed here, the reasons for hubs failing to be active could not be identified, probably because of the substantially higher complexity and heterogeneity of this network, including varying levels of clustering coefficients, as compared to the neuronal networks. For the latter, it was possible to verify that the correlation between activity and topology is enhanced if higher cumulative hierarchical instrengths were taken into account. With an artificially-constructed network, we demonstrated this should indeed be expected and found that higher clustering coefficients contribute to decreasing activity.

**Acknowledgements**

This work was supported by FAPESP and CNPq (Brazil).

# SUPLEMENTARY METHODS

*Cumulative Hierarchical Degrees*

The concepts of cumulative hierarchical in- and outdegree are defined for a given node $i$ taking into consideration its hierarchical level. The cumulative $1^{st}$ hierarchical indegree is the number of incoming edges extending from the immediate neighbors of node $i$ into that node, i.e. it corresponds to the traditional indegree of node $i$. These neighbors constitute the $1^{st}$ hierarchical level (see Supplementary Fig. 1). This number of connections between the hierarchical levels 0 and 1 plus those between levels 1 and 2 define the cumulative $2^{nd}$ level hierarchical indegree. More generally, the cumulative hierarchical indegree of node $i$ at hierarchy $h$ is equal to the number of directed edges extending from the nodes at hierarchy $h+1$ to the nodes at hierarchical level $h$, plus the indegrees obtained from hierarchy 1 to $h$-1. The cumulative hierarchical outdegrees are calculated similarly, but considering opposite edge directions. The cumulative hierarchical in- and outstrenghts are analogously defined, taking into account the weights of the edges (instead of only the number of edges).

# SUPPLEMENTARY FIGURES

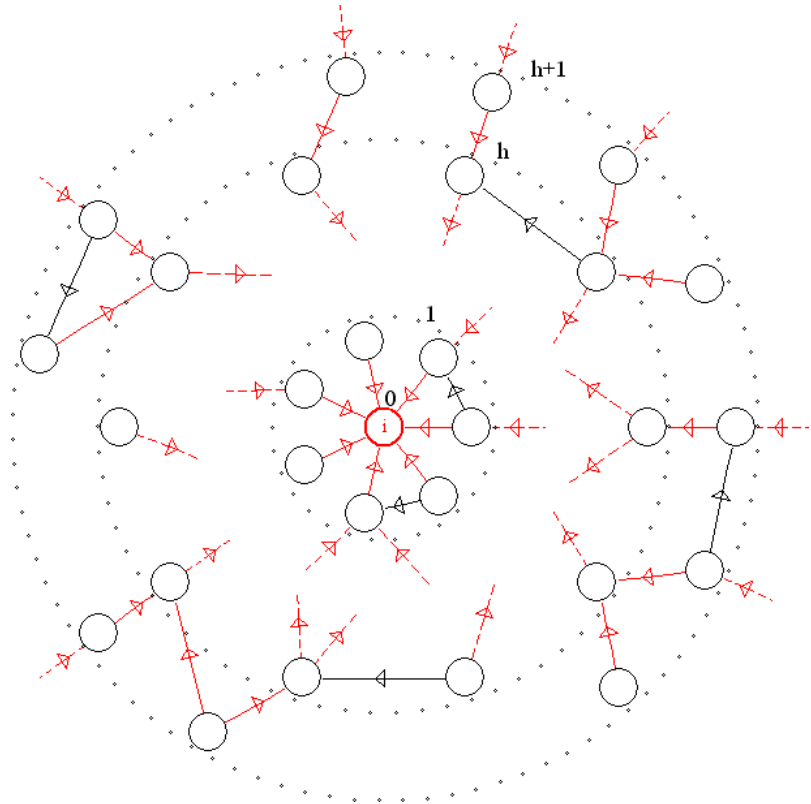

Supplementary Fig. 1 – Illustration of the concept of the hierarchical levels defined for a specific reference node *i*.

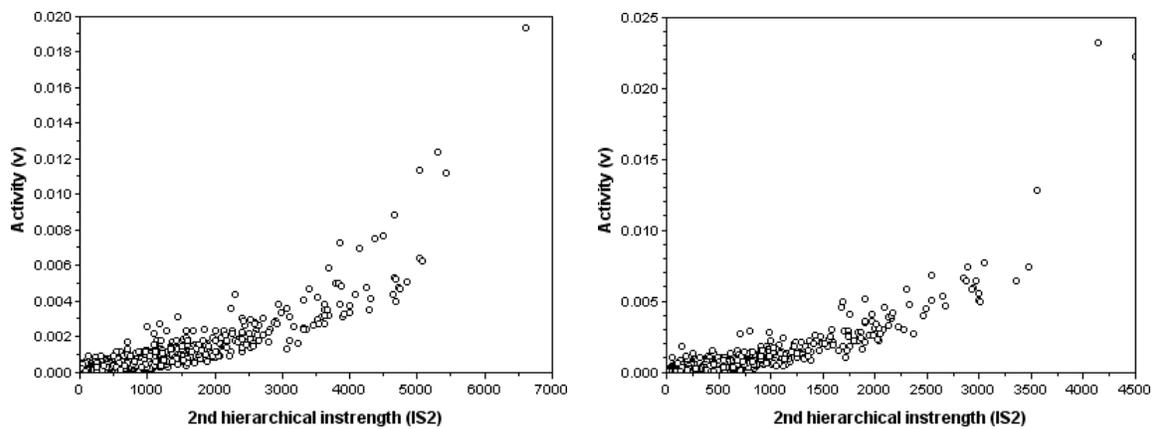

(a) (b)

Supplementary Fig. 2 – Activity in terms of the cumulative 2$^{nd}$ hierarchical instrength for the (a) Darwin network and the (b) Wodehouse network. The activity increases basically with the square of the 2$^{nd}$ cumulative hierarchical instrength.

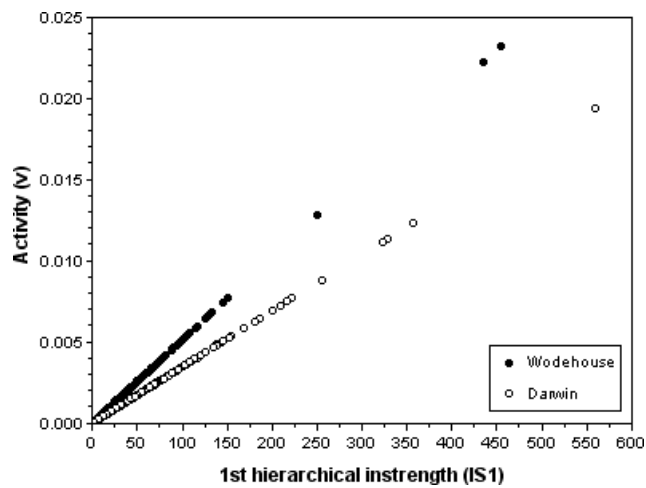

Supplementary Fig. 3 – Correlations between the frequency of visits to nodes and the traditional instrength for the two networks derived from the texts authored by C. Darwin and P.G. Wodehouse. The Wodehouse work is characterized by substantially steeper increase of frequency of visits.